# Temperature dependent screened electronic transport in gapped graphene

**Digish K Patel**[*,1]**, A C Sharma**[*,1] **and S S Z  Ashraf**[**,2]

[1] Physics Department, Faculty of Science, The M.S. University of Baroda, Vadodara-390002, Gujarat, India
[2] Physics Department, Faculty of Science, Aligarh Muslim University, Aligarh-202002, Uttar Pradesh, India

* e-mail jdiggish@gmail.com
* e-mail acs_phy@yahoo.com
** Corresponding author: e-mail ssz_ashraf@rediffmail.com

We report our theoretical calculations on the temperature and energy dependent electrical conductivity of gapped graphene within the framework of Boltzmann transport formalism. Since screening effects have known to be of vital importance in explaining the conductivity of gapless graphene therefore we first worked out the behaviour of the temperature dependent polarization function for gapped graphene as a function of wave vector and band gap, respectively. Polarization of gapped graphene has been compared with that of gapless graphene, bilayer graphene and 2DEG to see the effects of gap. It is found that the gapped graphene polarization function exhibits a strong dependence on temperature, wave vector and band gap and the effect translates to the conductivity of gapped graphene. The nature of conductivity in gapped graphene is observed to be non monotonic ranging from good to poor to semi conducting. We also find that the conductivity computed as a function of temperature by averaging over quasi-particle energy significantly differs from that computed at Fermi energy, suggesting that a notable contribution to temperature dependent conductivity is made by electrons close to the Fermi level.

## 1 Introduction

Graphene is a two dimensional novel material of many fascinating attributes. It is an intellectually stimulating object from both fundamental physics point of view and utility wise. The exceptional novel properties displayed by graphene make it a promising material for numerous applications in disparate fields [1]. One of the key features of single layer graphene (SLG) is its mass less Dirac type of low energy electron excitations having a linear energy dispersion relation; $\varepsilon_{sk} = shv_f|k|$, where  $v_f$ is the Fermi velocity - which is independent of electron density ($n_c$) in graphene and is about 300 times smaller than velocity of light in vacuum, and the subscript *sk* identifies the spin and wave vector of a state. The SLG is a gapless semiconductor with its valance and conduction bands touching at Dirac points. The absence of bandgap makes it challenging to create graphene-based transistors with large on/off ratios, which are required for logic applications. This instigated research on possible ways to open a gap between valance and conduction band in graphene for device applications.

Graphene systems which have extensively been studied are single layer gapless graphene (SLG), single layer gapped graphene (SGG) and bilayer graphene (BLG). Recent studies have demonstrated that a gap between valance band and conduction band can be opened in different ways, such as: (i) graphene placed on suitable substrate like silicon carbide (gap about 0.26 eV) [2-3] and boron nitride (gap about 53 meV) [4] , (ii) application of magnetic field to generate a dynamic gap [5], (iii) a small gap ($\sim 10^{-3}$ meV) is opened due to spin-orbit coupling or Rashba effect [6-7]. On opening of gap, the relativistic massless Dirac particle dispersion relation changes to $\mu_{sk} = \pm \sqrt{\varepsilon_{sk}^2 + \Delta^2}$ . The extra intrinsic mass, $\Delta = m v_f^2 = a\,\mu_f$ , is introduced due to a break in the graphene's sublattice symmetry. Here $\mu_f = \sqrt{\varepsilon_f^2 + \Delta^2}$ is the chemical potential at absolute zero temperature with $\varepsilon_f = \gamma k_f$ ($\gamma = 6.46$ eV Å is band parameter) and $a$, is normalised gap having values between 0 and 1 when Fermi level is considered to be lying above energy gap. In contrast to chiral linear energy dispersion of SLG, the BLG displays an energy dispersion that is chiral parabolic; $E_k = \hbar^2 k^2 / 2\,m^*$, where effective mass, $m^* = \gamma_1 / 2\,v_f^2$ , with interlayer hopping matrix element, $\gamma_1 = 0.39$ eV [8] and Fermi velocity $v_f \approx 1.1 \times 10^6$ m/s.

The charge transport in SLG, BLG and SGG display novel chirality that has attracted much theoretical and experimental attention [9-16]. Charge transport in graphene systems sharply differ from that of 2D electron gas (2DEG) observed in doped semiconductor heterostructures [15-16]. Though the energy dispersion in BLG is similar to that of 2DEG but the carrier density dependence of conductivity ($\sigma$) in BLG seems to manifest the same linear behaviour as that observed in SLG. Various theories of charge transport in graphene systems appear to suggest that considering a scattering mechanism based on screened charged impurities; one can obtain from the Boltzmann equation to give $\sigma$ that agrees with the experimental results. The intrinsic parameters that goven $\sigma$ and electron-impurity scattering rate ($\hbar/\tau$) in SLG, BLG and SGG are quasi particle energy ($\varepsilon$), temperature (T) and carrier concentration ($n_c$), and additionally energy gap ($\Delta$) in BLG and SGG. Theoretical understanding of $\sigma$ requires detailed investigations on how polarization function and $\sigma$ depends on $q$, T and $\Delta$. There have been several calculations of the polarization function and its properties for SLG & BLG [11-12, 15-20]. Also the conductivity as a function of various influencing parameters for SLG and BLG has been reported in several publications [9-16, 15-17].

In this paper, we report our theoretical investigations on SGG polarization function and SGG conductivity and its dependence on various governing intrinsic parameters. The conductivity in SGG has been dealt within the Boltzmann transport theoretical approach. Most of the existing theoretical studies on SGG have been performed at zero temperature by calculating static and dynamical polarization functions at zero temperature and zero magnetic field [21-28]. It has been shown that the screening of charged impurities at large distances in SGG differs from that in SLG by slower decaying of Friedel oscillations ($1/r^2$ instead of $1/r^3$), similar to that observed in conventional 2D system. As implied above, the main motive of our work has been to investigate how the polarization function of SGG and hence the conductivity of gapped graphene varies with temperature and gap. A strong temperature dependent conductivity has been reported in high mobility SLG samples [15- and references there in]. Also, at very low temperatures, the Fermi function behaves like step function and hence there is no significant difference between conductivities computed by averaging over $\varepsilon$ and at Fermi energy. However, as temperature increases, conductivity computed by averaging over $\varepsilon$ significantly differs from that computed at Fermi energy. The conductivity computed by averaging over $\varepsilon$ and not the conductivity at Fermi energy should be compared with experimental results.

The paper is organized as follows: Section 2 briefly discusses the formalism and section 3 contains a detailed discussion on temperature dependent polarization function of SGG. In section 4 we report the results on conductivity of SGG as a function of temperature, energy bandgap, carrier density and coupling constant with their discussion, and finally we conclude our findings in section 5.

## 2 Essential formalism

As outlined in the introduction, in the calculation of conductivity at $T = 0\ K$, all the charge carriers are at the Fermi level and the averaging of the relaxation time can be ignored. At finite temperature the carrier concentration is expressed in terms of distribution function therefore the relaxation time should be taken as the average of the relaxation time of individual charge carriers. Therefore it is not appropriate to ignore the averaging, since $\boldsymbol{\tau}$ at Fermi level can significantly differ from $\langle\boldsymbol{\tau}\rangle$ particularly at moderate and high temperatures. In terms of the average relaxation time $\langle\boldsymbol{\tau}\rangle$, the temperature dependent conductivity $\boldsymbol{\sigma(T)}$ for SGG within the Boltzmann transport formalism can be expressed as [9];

$$\sigma(\mathrm{T}) = (2e^2/h)\left(\varepsilon_f\,\langle\tau\rangle/\hbar\right) \tag{1}$$

where $\langle\tau\rangle$ is the average value of $\tau$ over all possible values of quasiparticle energy, $\varepsilon_{sk}$ . The energy averaged finite temperature scattering time is given by [9],

$$\langle\tau\rangle = \int \varepsilon_{sk}\,\tau(\varepsilon_{sk},\mathrm{T})\left(-\frac{\partial f}{\partial \varepsilon_{sk}}\right)\mathrm{d}\varepsilon_{sk}/\int \varepsilon_k\left(-\frac{\partial f}{\partial \varepsilon_{sk}}\right)\mathrm{d}\varepsilon_{sk}. \tag{2}$$

Where $f(\varepsilon_{sk}) = 1/\left(1 + e^{\left((\varepsilon_{sk}-\mu_f)\beta\right)}\right)$ is the Fermi distribution function, $\tau(\varepsilon_{sk},\mathrm{T})$ is the finite temperature and energy dependent scattering time of an electron scattered by disorder or statically screened Coulomb potential, given by [15]

$$\frac{\hbar}{\tau(\varepsilon_{sk},\mathrm{T})} = 2\pi n_i \int \frac{\mathrm{d}k'}{(2\pi)^2}\langle|V_{sk'\,sk}|^2\rangle(1 - \cos\theta_{kk'})\,\delta\left(\sqrt{\varepsilon_{sk}^2 + \Delta^2} - \sqrt{\varepsilon_{sk'}'^2 + \Delta^2}\right). \tag{3}$$

Where $n_i$ is the concentration of impurity centers, $\theta_{kk'}$ is the scattering angle between the scattering in and out wave vectors $k$ and $k'$, $\langle|V_{sk'\,sk}|^2\rangle$ is the norm square of matrix element of scattering potential averaged over configurations of impurities expressed in terms of charge impurity potential,

$$\langle|V_{sk'\,sk}|^2\rangle = \left(\frac{V(q)}{\epsilon(q,T)}\right)^2\left(\left(\frac{\varepsilon_{sk}^2 + 2\Delta^2}{\varepsilon_{sk}^2 + \Delta^2}\right) + \left(\frac{\varepsilon_{sk}^2}{\varepsilon_{sk}^2 + \Delta^2}\right)\cos\theta\right). \tag{4}$$

Where, $V(q) = 2\pi e^2/\kappa q$ is the two dimensional Fourier transform of the bare charge impurity Coulomb potential, where $e$ is the electronic charge, $\kappa$ is the average background dielectric constant ($\approx 5.5$ for graphene placed on SiC with other side being exposed to air), $q = |k - k'| = 2\,(\varepsilon_{sk}/\gamma)\,\sin(\theta/2)$ is the momentum transferred to a scattered electron, $\epsilon(q,T)$ is the temperature dependent static dielectric function, which within linear response theory is given by

$$\epsilon(q,T) = 1 + V(q)\Pi(q,T). \tag{5}$$

Substituting Eqs. (4) & (5) into Eq. (3), and then simplifying the integrals, we obtain

$$\frac{\hbar}{\tau(\varepsilon_{sk},T)} = 4n_i D(\varepsilon_{sk}) \int_0^\pi d\theta \left(\frac{\pi e^2}{\kappa q\,\epsilon(q,T)}\right)^2 (1 - \cos\theta)\,\left(\left(\frac{\varepsilon_{sk}^2 + 2\Delta^2}{\varepsilon_{sk}^2 + \Delta^2}\right) + \left(\frac{\varepsilon_{sk}^2}{\varepsilon_{sk}^2 + \Delta^2}\right)\cos\theta\,\right). \tag{6}$$

Where, the factor $(1 - \cos\theta)$ which is invariably present in the Boltzmann transport equation weighs the amount of backward scattering of the electrons by impurity. In case of SLG, factor $\left(\left(\frac{\varepsilon_{sk}^2 + 2\Delta^2}{\varepsilon_{sk}^2 + \Delta^2}\right) + \left(\frac{\varepsilon_{sk}^2}{\varepsilon_{sk}^2 + \Delta^2}\right)\cos\theta\,\right)$ reduces to $(1 + \cos\theta)$ that suppresses the large angle scattering in SLG. However, in case of SGG there is breaking of the sub lattice symmetry because of the bandgap, which contributes to the large angle scattering with increasing band gap. At lower temperatures, the derivative of Fermi distribution function, $\partial f/\partial \varepsilon_k$ behaves like step function and therefore it is assumed that for all practical purposes there should not be any significant difference in temperature dependent conductivity calculated at Fermi energy or with the use of Eq. (1).

Rest of the paper, we used following scaled parameters: $x = \varepsilon/\varepsilon_f$, $z = q/k_f$, $y = T/T_f$, $a = \Delta/\mu_f$, $\alpha = 4e^2/\gamma\kappa$ (dimensionless coupling constant, which has value $\approx 1$ (& 2) for graphene sheets on SiC (BN) substrate) and $N = n_c/n_i$ (the scaled carrier concentration). The normalized conductivity at finite T for SGG is then obtained from the ratio; $\sigma(y)/\sigma(y = 0) = \tau(y)/\tau(y = 0)$, where $\tau(y = 0)$ has been computed by taking $y = 0$ in Eq. (6).

## 3 Temperature dependent polarizability of gapped graphene

The polarization function is an important quantity in calculating scattering rate of screened electron gas. The polarization function for the cases of both doped and undoped gapped graphene has been calculated at zero temperature [24-25, 27] as well as at finite temperature [21] in the recent past and the effect of band gap on the ground state properties of Dirac electrons in a doped graphene at zero temperature has been studied [24]. We present here the detailed analysis of the numerically computed temperature dependent polarization function for SGG and its comparison with that of SLG, BLG and 2DEG. Earlier, analytical results only in the asymptotic limits have been presented and detailed analysis of the polarization function at all $q$-values as well as temperature and gap values is missing [21]. The temperature dependent polarization function for SGG in the random phase approximation (RPA) is given by [21],

$$\Pi(q,\mathrm{T}) = \Pi_{vacuum}(q,\mathrm{T}) + \Pi_{matter}(q,\mathrm{T}). \tag{7}$$

Where,

$$\frac{\Pi_{vacuum}(q,\text{T})}{D_0} = \frac{\Delta}{2\,\mu_f} + \frac{q\,\gamma}{4\,\mu_f}\left(1 - \frac{4\,\Delta^2}{\gamma^2 q^2}\right)\tan^{-1}\left(\frac{q\,\gamma}{2\,\Delta}\right)$$

(8)

and

$$\frac{\Pi_{matter}(q,\text{T})}{D_0} = \frac{\mu_T}{2\,\mu_f} - \frac{\Delta}{\mu_f} + \frac{1}{\beta}\left(\ln\left[1 + e^{-\beta(\mu_T - \Delta)}\right] + \Delta \to -\Delta\right) - \frac{\gamma^2}{q\,\mu_f}\left\{\int_0^{\frac{q}{2}}\frac{k\,dk}{\sqrt{\gamma^2 k^2 + \Delta^2}}\right.$$

$$\left.\times\left(\sqrt{\frac{q^2\gamma^2}{4} - \gamma^2 k^2} - \frac{2\Delta^2}{\sqrt{\frac{q^2\gamma^2}{4} - \gamma^2 k^2}}\right)\left(\frac{1}{1 + e^{\beta\left(\sqrt{\gamma^2 k^2 + \Delta^2} - \mu_T\right)}} + \mu_T \to -\mu_T\right)\right\}.$$

(9)

Here $\mu_T$ is the finite temperature chemical potential determined by the conservation of the total electron density, defined by

$$\left(\frac{T_f}{T}\right)^2 = 2(F_1(\beta\,\mu_T) - F_1(-\beta\,\mu_T)).$$

(10)

Where $\beta = 1/k_b T$, and $F_1(\beta\,\mu_T)$ is given as;

$$F_1(\beta\,\mu_T) = \int_0^\infty \frac{t}{1 + e^{(t - \beta\,\mu_T)}}\,dt.$$

(11)

With the use of the scaled variables define in section 2, Eq. (7) using Eqs. (8) & (9) takes the form;

$$\Pi(z,y,a) = \Gamma - \frac{a}{2} + y\left(\ln\left[1 + e^{-\frac{(\Gamma - a)}{y}}\right] + a \to -a\right) + \frac{z\sqrt{1 - a^2}}{4}\left(1 - \frac{4a^2}{z^2(1 - a^2)}\right)\tan^{-1}\left(\frac{z\sqrt{1 - a^2}}{2a}\right) - (1 - a^2)$$

$$\times\left\{\int_0^{\frac{z}{2}}\frac{x\,dx}{\sqrt{x^2(1 - a^2) + a^2}}\left(\sqrt{1 - \frac{4x^2}{z^2}} - \frac{4a^2}{z^2(1 - a^2)\sqrt{1 - \frac{4x^2}{z^2}}}\right)\left(\frac{1}{1 + e^{\left(\sqrt{x^2(1 - a^2) + a^2} - \Gamma\right)/y}} + \Gamma \to -\Gamma\right)\right\},$$

(12)

where $\Gamma = \mu_T/\mu_f$. The above equation yields the prior reported results [15, 17, 22] for polarizability at T = 0 for SGG and at finite temperature for SLG when $a = 0$. The computed normalized polarizability as a function of $z$ from Eq. (12) is plotted in Fig. 1(a) at T = 0 for different gap values and in Fig. 1(b) at $a = 0$ for different temperature values. As can be seen from Fig. 1(a), the effect of introducing gap in electronic spectrum at T = 0 is almost unnoticeable for $q < 2k_f$, which means that the intraband and interband transitions almost cancel at zero temperature and the total polarizability is static, similar to that in SLG [18]. But when $q > 2k_f$ (large momentum transfer regime); (i) the magnitude of polarizability versus wave vector curve decreases on increasing the gap, at all $q$-values, and (ii) for $a > 0.6$ the behaviour of polarizability versus wave vector curve for SGG resembles to a great extent with that of 2DEG polarizability which is in stark contrast to SLG where the polarizability increases for $q > 2k_f$. This means that in SGG the interband transitions dominate over the intraband transition for large wavevectors, suggesting that the scattering by the screened coulomb potential is much reduced due to enhanced screening in this limit. This also implies that for $a = 0$, polarizability (SGG) shows relativistic characteristics while at $a \approx 1$ it reflects the non-relativistic nature of 2DEG caused by breaking of sub-lattice symmetry. Unlike the case of T = 0 and $a \neq 0$ where the polarizability stays constant for $q < 2k_f$, the polarizability shows a monotonically increasing behaviour for T $\neq$ 0 at $a = 0$ for $q$ values even in the range of $0 < q < 2k_f$, similar to the behaviour of SLG as is evident from Fig. 1(b) and also corroborated from Fig. 1(h). Also, the magnitude of polarizability versus wave vector curve enhances on increasing values of T. This is due to the enhanced electronic transitions to the conduction band from valance band with the increase of temperature. Variation of polarizability with wave vector at larger gap value $a(= 0.9\ \&\ 0.99)$ and at low temperatures are shown in Figs. 1(c) & 1(d). The sharp decline seen in polarizability from a constant value at $z = 2$ and T = 0, changes to a smooth variation on increasing the temperature. This abrupt decline in polarizability at $z = 2$ is associated with Friedel oscillations, which at finite and reasonably higher temperatures wash out.

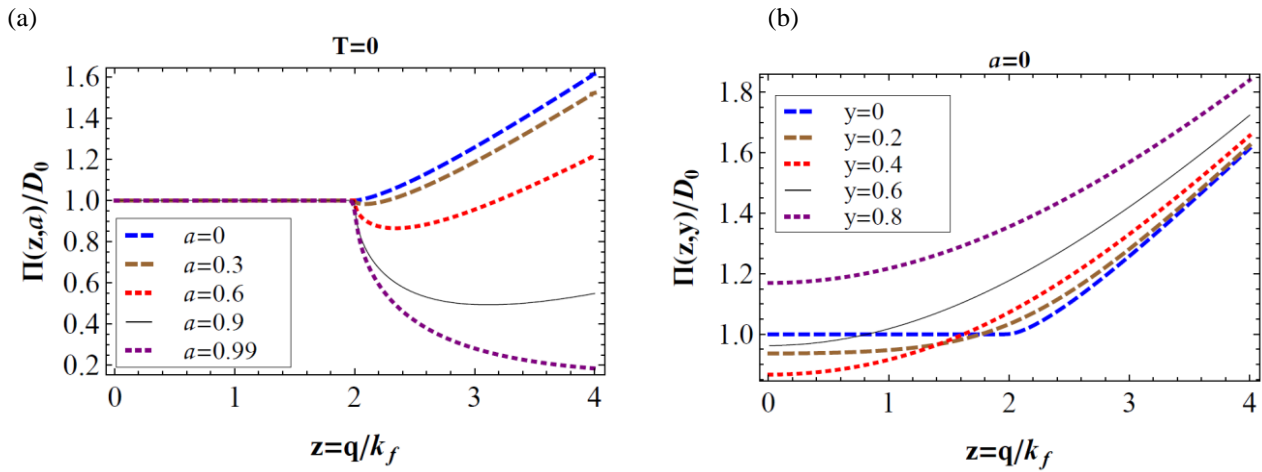

(a) T=0 — $\Pi(z,a)/D_0$ vs $z = q/k_f$: curves for $a=0$, $a=0.3$, $a=0.6$, $a=0.9$, $a=0.99$.

(b) $a=0$ — $\Pi(z,y)/D_0$ vs $z = q/k_f$: curves for $y=0$, $y=0.2$, $y=0.4$, $y=0.6$, $y=0.8$.

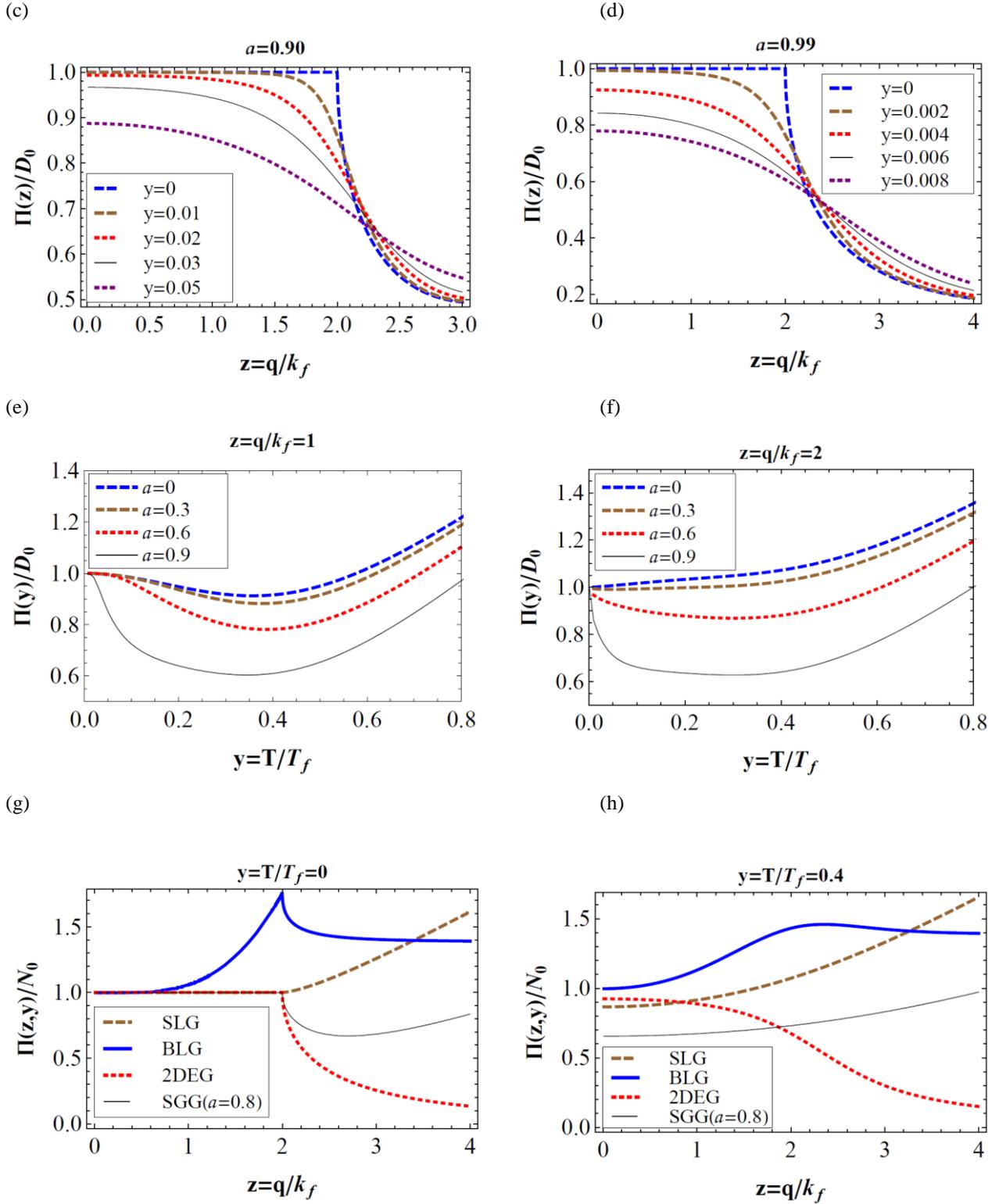

**Figure 1** Wave vector dependent polarizability (a) for different values of gap at $T = 0$ and (b) for different values of temperature at $a = 0$. Figures (c) and (d) show the wave-vector dependence of polarizability for different values of tem-

perature at gap values of $a = 0.9$, and 0.99 respectively. The curves in this case are similar to polarization function for a Si(001) inversion layer with $2 \times 10^{12}$ electron per cm$^2$ [29]. Temperature dependent polarizability for different gap values at; (e) $z = 1$ and (f) $z = 2$. Comparative plot of polarization function of SLG, BLG, SGG & 2DEG at zero and finite temperature ($y = 0.4$) are shown in Figs. 1(g) and 1(h), respectively. Here $D_0 = 2\mu_f/(\pi\gamma)$.

The computed polarizability as a function of temperature using Eq. (12) for different values of $a(= 0.0, 0.3, 0.6 \& 0.9)$ at two values of $z(= 1 \& 2)$ are plotted in Figs. 1(e) & 1(f). A slight change (decrease) in nature of polarizability versus wave vector curve can be noticed on enhancing $z$ from 1 to 2. Also, as can be noticed from the figures, the polarizability first declines with temperature and after hitting a minimum increases almost linearly for all nonzero values of $a$. A reverse trend observed at higher temperatures compared to that at low temperature regime is indicative of phase transition taking place in SGG. Finally in Figs. 1(g) & 1(h) the comparative plot of the polarization function of SLG, SGG, BLG & 2DEG are shown at zero and finite temperature ($y = 0.4$), respectively. The interplay of linear energy band dispersion relation, chirality, bandgap and temperature endow SGG with overall strange screening properties which are a mixture of SLG, BLG and 2DEG screening properties. It is known that SLG also exhibits strange screening properties which arise because of a combination of metallic screening due to intraband transitions and insulating screening due to interband transitions; that all ultimately stems from the chiral relativistic dispersion relation [18].

## 4 Temperature-dependent Conductivity of gapped graphene

This section reports our results on energy averaged conductivity as a function of temperature calculated using Eq. (1), which is then compared with that calculated at Fermi energy. It is found that when we incorporate the temperature dependence in dielectric function formalism, we observe significant difference between these two conductivities as function of temperatures. We thus find that it is grossly misleading to calculate temperature dependent conductivity at Fermi energy for comparison with experimental results. Our computed numerical results on normalized conductivity as a function of temperature, with the use of $1/\langle\tau\rangle$ and $1/\tau_{\varepsilon_f}$, are plotted in Figs. 2(a) to 2(d) for different values of $a$. The figures clearly demonstrate that the difference between two values of conductivity is insignificant only for temperatures very close to zero and the difference grows with increasing temperatures. For $a(= 0, 0.3 \& 0.6)$, $\sigma(y)/\sigma(y = 0)$ using $1/\langle\tau\rangle$ initially remains almost constant and then it increases with $y$, which is not the case when $1/\tau_{\varepsilon_f}$ used to compute $\sigma(y)/\sigma(y = 0)$. A phase transition on changing temperature is also indicated by conductivity computed using energy averaged scattering rate for higher values of $a > 0.6$, as can be seen from Figs. 2(a) to 2(d). Elaborated behaviour of $y$ verses $\sigma(y)/\sigma(y = 0)$ near the transition point is shown insets of figures. The green dot of curve indicates turning points of set curves.

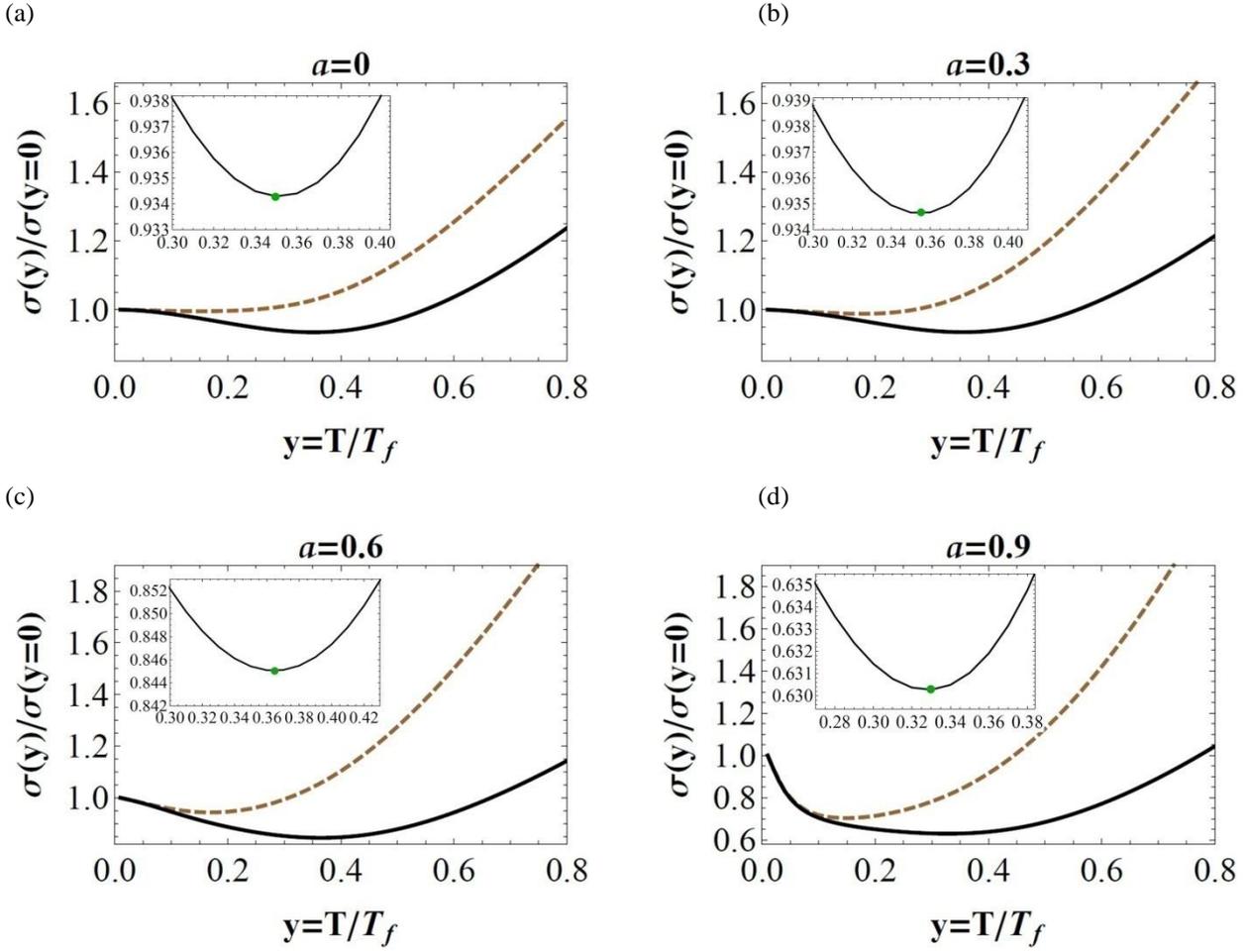

**Figure 2** Conductivity as a function of temperature calculated at Fermi energy (solid black line) and at average energy (dashed brown line) for different gap values; (a) $a = 0$, (b) $a = 0.3$, (c) $a = 0.6$ and (d) $a = 0.9$. Here we used coupling constant, $\alpha = 1$ (for SiC). The green spot in insets show minimum value of conductivity.

Change in nature of conduction, from poor conductor to semiconductor, takes place when $y \gtrsim 0.35$ for $a(= 0, 0.3, 0.6$ & $0.9)$. Strikingly opposite behaviour of conductivity with temperature in low and high temperature regimes is indicative of phase transition that can be obtained by selecting appropriate values of $a$ and $y$ in SGG. Curves exhibit a minimum at $y \approx 0.35$ and the change from poor metallic nature for $a(= 0, 0.3$ & $0.6)$ to good conductor nature at $a = 0.9$ in low temperature regime. The behaviour of the conductivity can be understood as follows; the increase in band gap reduces the conductivity in low temperature regime due to electron has insufficient excitation energy to jump from valance band to conduction band but increase in temperature increases the excitations and hence the raises the magnitude of conductivity. The Fig. 3(a) showing the variation of conductivity with bandgap at different temperatures, corroborates the above behaviour. In Fig. 3(b) shows the variation of conductivity as a function of temperature for $\alpha = 2$ which corresponds to graphene on BN (Boron Nitride) substrate. The higher values of $\alpha$ indicate a strong coupling in terms of elec-tron-electron interaction. A slight change (decrease) in nature of conductivity versus temperature curve can be noticed on enhancing coupling constant, $\alpha$, from 1 to 2.

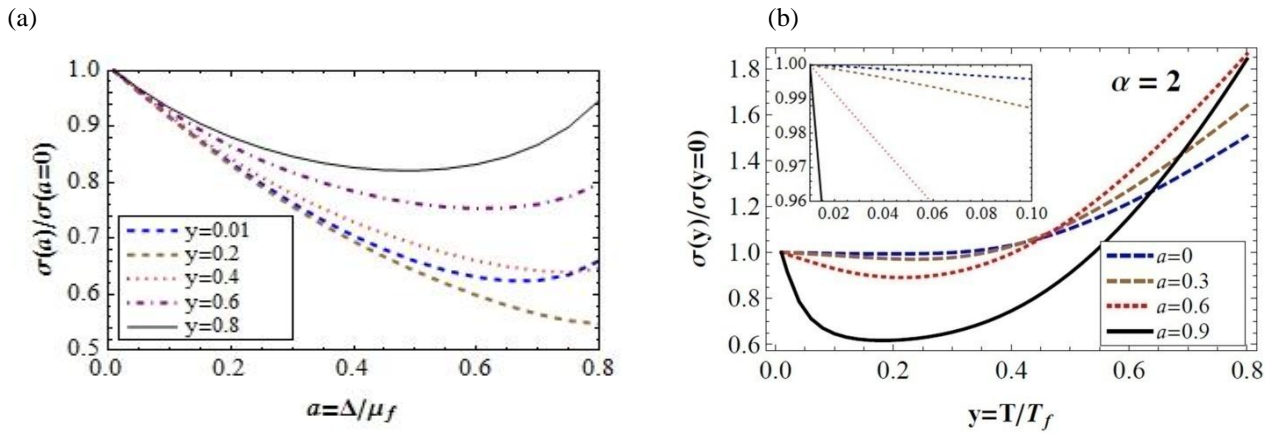

**Figure 3** Conductivity (a) as a function of band gap for different values of temperature $y(= 0, 0.2, 0.4, 0.6 \ \& \ 0.8)$, using $\alpha = 1$, (b) as a function of temperature calculated at average energy for different gap values gap $a(= 0, 0.3, 0.6 \ \& \ 0.9)$ with coupling constant, $\alpha = 2$.

## 5 Conductivity as a function of carrier concentration

Our computed conductivity using Eq. (1) as a function of carrier concentration, $n_c$ at different temperatures is plotted in Figs. 4(a) & 4(b) for two values of $a(= 0 \ \& \ 0.6)$. Almost a linear enhancement of conductivity with $n_c$ is seen for both values of $a$, similar to the experimentally observed behaviour in case of SLG [9]. Also the magnitude of conductivity increases with the increase in temperature. The Figs. 4(c) & 4(d) depict the effect of variation of conductivity with band gap at $y = 0.01$ and $y = 0.8$, respectively, for three values of carrier concentration $N(= 0.5, 1.0 \ \& \ 2.0)$. As can be noticed from the figure; this reconfirms that the enhancement in the carrier concentration and temperature increases the conductivity whereas the increase in band gap reduces the conductivity.

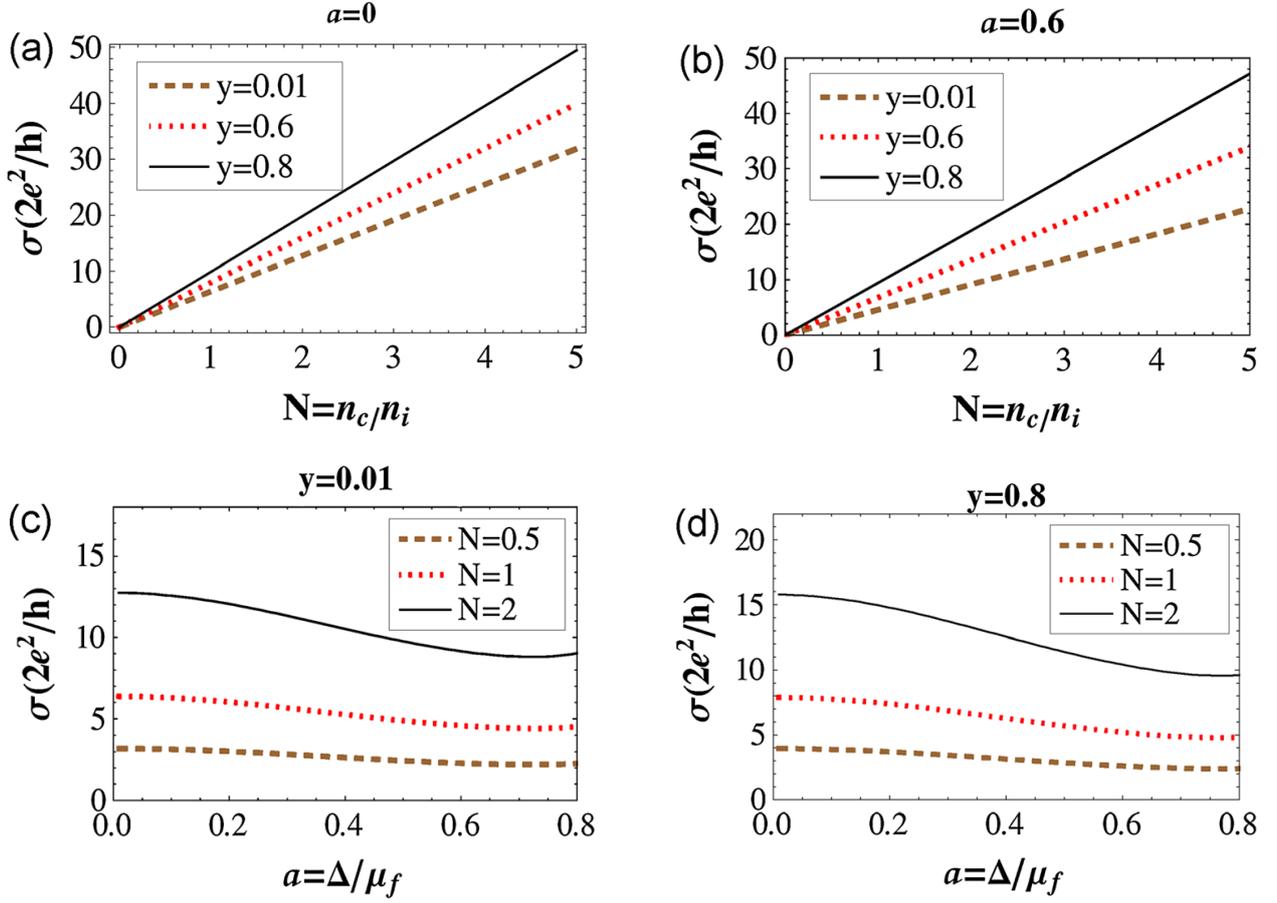

**Figure 4** Conductivity as (i) a function of $n_c/n_i$ for different temperatures of $y(=0, 0.6 \& 0.8)$ with gap values; (a) $a = 0$, (b) $a = 0.6$, (ii) a function of bandgap for $N(= 0.5, 1 \& 2)$ at (c) $y = 0.01$ and (d) $y = 0.8$.

## 6 Conclusions

We computed temperature dependent polarization function and the conductivity of doped gapped graphene considering electron-impurity scattering as the dominant source and neglecting all other scattering phenomena (e.g. electron-phonon, electron-electron), within the Boltzmann transport theory. The interplay of linear energy band dispersion relation, chirality, bandgap and temperature endow SGG with overall strange screening properties which are a mixture of SLG, BLG and 2DEG screening properties. We find that the nature of conduction in gapped graphene changes from good to poor semiconducting on varying values of $y$ and $a$. Our results show that $y \leq 0.4$ conductivity behaves like that of poor metal when $0 \leq a \leq 0.6$ and very good conducting metallic behaviour when $a \geq 0.9$, whereas it displays an insulating behaviour at higher temperatures ($y > 0.4$) for both gapless and gapped graphene. Metallic and semiconducting behaviour at low temperatures and high temperatures is indicative of phase transition that can occur by selecting appropriate values of $a$ and $y$ in SGG. We also notice that the metallic nature can be enhanced by increasing the coupling constant value. We also obtained numerical results of $T$-dependent conductivity as a function of carrier concentration which shows linear behaviour as observed experimentally, and also shows an increase in magnitude with the increase in temperature and decrease in magnitude with the increase in bandgap. Our results on conductivity of gapped graphene are of significance as any experimental work on graphene begins with a characterization of its electrical conductivity.